# An Improved DFT-based Channel Estimation for Mobile Communication Systems

*H. Yu and C. Yang*

*School of Electronic and Optical Engineering, Nanjing University of Science and Technology*

*Abstract*—Channel estimation is one of the most important parts in current mobile communication systems. Among the huge contributions in channel estimation studies, the discrete Fourier transform (DFT)-based channel estimation has attracted lots of interests since it can not only be easily implemented but also have acceptable performance in practical systems. In this paper, we propose an improved DFT-based channel estimation scheme that tries to clean the reference signals in the time domain before being used for interpolation using the estimated noise variance from reference signals on multiple orthogonal frequency division multiplexing (OFDM) symbols via the property of DFT. The proposed channel estimation scheme does not need any *priori* channel information. We validate the proposed channel estimation scheme in various channel models via simulations and comparison with conventional channel estimation schemes.

## I. INTRODUCTION

WITH the widespread of various applications on smart devices, the demand for high data rate in mobile communication systems is growing explosively. Orthogonal frequency division multiplexing (OFDM) has been widely adopted in the fourth generation (4G) mobile communication systems for its tremendous advantages not only in combating multi-path fading but also in parallel data transmission to obtain high data rate using a simple one-tap equalizer [1].

It has been well known that the use of differential phase-shift keying (DPSK) modulation in OFDM systems would avoid the tracking of a time varying channel (or no need for channel estimation). However, this scheme will limit the number of bits per symbol and result in a 3 dB loss in signal-to-noise ratio (SNR) [2]. Therefore, higher modulation schemes, for example, 256QAM, cannot be employed in this scheme to obtain higher data rate.

In today's mobile communication systems, coherent demodulation of received signal plays an important role in achieving high data rate. And channel estimation (CE) is a necessary part in realizing coherent demodulation. In channel estimation, reference signals are inserted into the data at the transmitter and used at the receiver. There are two types of reference signal patterns, i.e., block-type reference signal pattern and comb-type reference signal pattern. In block-type reference signal pattern, reference signals are inserted at the head of the data in a transmission. Then, at a receiver, the estimated channels at the reference signals are used to demodulate the receive data signals that follow the reference signals. Therefore, block-type reference signal pattern is effective in low mobility environment where the channel does not vary significantly in a transmission. In comb-type reference signal pattern, reference signals are evenly inserted into the data. At the receiver, the estimated channels at the reference signals are used to interpolate the channels where data are transmitted on. As a result, comb-type reference signal pattern has better channel tracking capability than block-type reference signal pattern, particularly in frequency-time selective channel conditions, though it's realized at the cost of a higher complexity channel estimation algorithm (e.g., the need for interpolation). Therefore, we focus on the comb-type reference signal pattern in this paper.

Among the fruitful results in channel estimation studies that use comb-type reference signal pattern, the discrete Fourier transform (DFT)-based channel estimation achieves a good tradeoff in complexity and performance. Therefore, it has been extensively studied in cellular communication systems [3]-[6]. In [3], the DFT-based channel estimation was first proposed for OFDM communication systems under time-varying channels. Then, the performance of three low-complexity DFT-based channel estimators for OFDM systems was analyzed, and the analytical expressions for mean squared error (MSE) were presented in [4]. Their analysis shows that this type of estimators may experience an irreducible error floor at high SNRs. In [5], a DFT-based channel estimation scheme for OFDM systems in non-sample-spaced multipath channels was proposed. They use a symmetric extension DFT method to reduce the leakage power efficiently thus significantly eliminate the error floor. In [6], the DFT-based channel estimator for long term evolution (LTE)-Advanced uplink was investigated and a frequency-domain weighting scheme was proposed to improve the degraded performance due to the non-orthogonal channels of multiple antennas of the current reference signal pattern at the time domain. Their scheme does not rely on the channel covariance matrix, and the weights can be generated offline.

In this paper, we propose an improved DFT-based channel estimation scheme that tries to make the reference signals as accurate as possible before being used for interpolation. Specifically, reference signals on multiple consecutive OFDM symbols are firstly used to estimate noise variance via the property of DFT. Then the estimated noise variance is used to clean the reference signals which are transformed into time domain channel impulse responses. Compared to conventional DFT-based channel estimation schemes, no *priori* channel information is need which makes it suitable for a great variety of channel models.

The remainder of this paper is as follows. In Section II, we describe the system model of an OFDM system and the conventional DFT-based channel estimation scheme. Then an improved DFT-based channel estimation scheme is proposed in Section III. And the simulation results are provided to validate the effectiveness and superiority of the proposed scheme in Section IV. Conclusions are given in Section V.

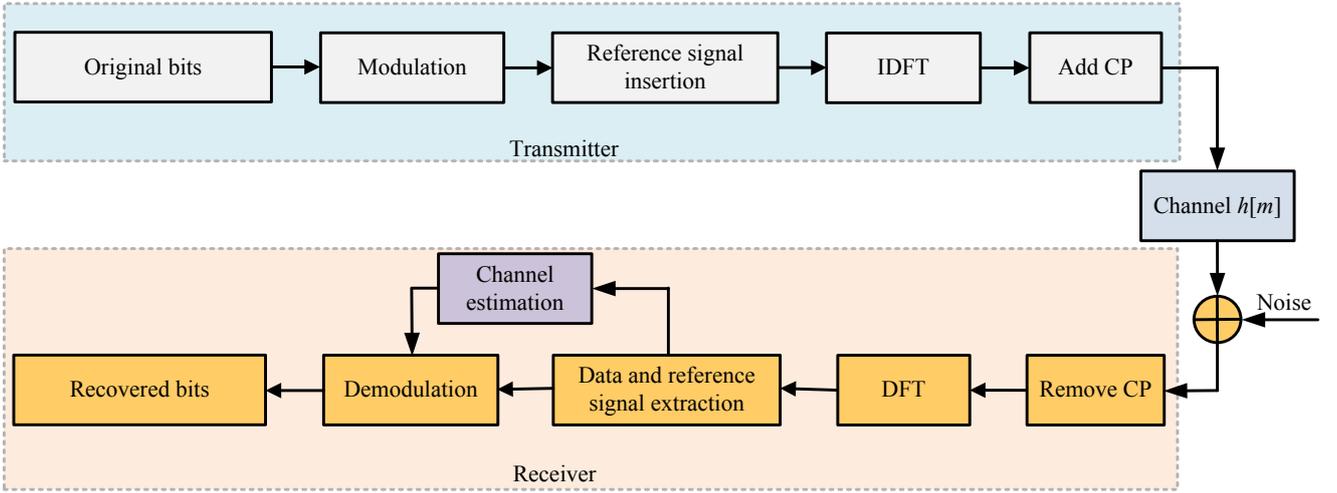

Fig. 1. The signal processing procedure in an OFDM system.

## II. THE SYSTEM MODEL AND CONVENTIONAL DFT-BASED CHANNEL ESTIMATION

In this section, we give a brief description of the system model of an OFDM system and the conventional DFT-based channel estimation scheme. The signal processing procedure in an OFDM system is depicted in Fig. 1. In Fig. 1, at the transmitter, the original information bits are firstly modulated into complex data symbols, and then reference signals for channel estimation are inserted into the complex data symbols. An IDFT operation is applied to the complex data symbols and reference signals to get the time domain signals. To avoid multi-path interference and inter-OFDM symbol interference, cyclic prefix (CP) is added in the time domain. The transmitted data and reference signals may be corrupted by the wireless channel and noise. At the receiver, the CP is firstly removed and a DFT operation is performed to the received signals. Then the reference signals are extracted to perform the channel estimation, and the estimated channel is used for demodulation of the extracted data. Finally, the original information bits are recovered via demodulation.

In the OFDM system, we assume there are $N$ subcarriers ($N_p$ of which are reference signal subcarriers used for channel estimation) on each OFDM symbol. With perfect frequency and time synchronization, the $n$th received baseband reference signal on the $m$th OFDM symbol can be written as follows,

$$Y_p[n,m] = H_p[n,m]X_p[n,m] + W_p[n,m], \quad (1)$$

where $0 \leq n \leq N_p - 1, 0 \leq m \leq M-1$, $M$ is the number of OFDM symbols in a subframe or a block. $X_p[n,m]$ is the $n$th reference signal of the $m$th OFDM symbol and satisfies $X_p[n,m]X_p^*[n,m] = 1$. $H_p[n,m]$ is the channel corresponding to the $n$th reference signal subcarrier of the $m$th OFDM symbol and is modeled by a complex Gaussian variable with zero-mean and unit-variance. $W_p[n,m]$ is the noise and is distributed according to $\mathcal{CN}(0,\sigma^2)$. And $A^*$ denotes the conjugate of $A$.

The channel corresponding to the $n$th reference signal subcarrier of the $m$th OFDM symbol is obtained via least square (LS) estimation using the received reference signals in (1), i.e.,

$$\begin{aligned}\hat{H}_p[n,m] &= Y_p[n,m]X_p^*[n,m] \\ &= H_p[n,m]X_p[n,m]X_p^*[n,m] + W_p[n,m]X_p^*[n,m] \quad (2)\\ &= H_p[n,m] + W_p[n,m]X_p^*[n,m],\end{aligned}$$

where $0 \leq n \leq N_p - 1, 0 \leq m \leq M-1$, and $W_p[n,m]X_p^*[n,m]$ is also distributed according to $\mathcal{CN}(0,\sigma^2)$.

For the simplicity of notations, let's arrange the LS-estimated channels corresponding to the reference signal subcarriers in (2) into a matrix as follows,

$$\hat{\mathbf{H}}_p = \begin{bmatrix} \hat{H}_p[0,0] & \cdots & \hat{H}_p[0,m] & \cdots & \hat{H}_p[0,M-1] \\ \vdots & \ddots & \vdots & \ddots & \vdots \\ \hat{H}_p[n,0] & \cdots & \hat{H}_p[n,m] & \cdots & \hat{H}_p[n,M-1] \\ \vdots & \ddots & \vdots & \ddots & \vdots \\ \hat{H}_p[N_p-1,0] & \cdots & \hat{H}_p[N_p-1,m] & \cdots & \hat{H}_p[N_p-1,M-1] \end{bmatrix} \quad (3)$$

$$= \begin{bmatrix} \hat{\mathbf{H}}_p[0] & \cdots & \hat{\mathbf{H}}_p[m] & \cdots & \hat{\mathbf{H}}_p[M-1] \end{bmatrix},$$

where $\hat{\mathbf{H}}_p[m]$ is the $m$th column of matrix $\hat{\mathbf{H}}_p$, which is the channels corresponding to the reference signal subcarriers on the $m$th OFDM symbol.

In conventional DFT-based channel estimation schemes [7], the estimation of the channels on non-reference signal subcarriers is performed on each OFDM symbol using the LS-estimated channels on reference signal subcarriers. For example, on the $m$th OFDM symbol, an $N_p$-point IDFT is firstly employed to transform the frequency domain channels $\hat{\mathbf{H}}_p[m]$ in (3) into the time domain channel impulse response, i.e.,

$$\begin{bmatrix} \hat{h}_p[0] & \cdots & \hat{h}_p[i] & \cdots & \hat{h}_p[N_p-1] \end{bmatrix}^T = IDFT\{\hat{\mathbf{H}}_p[m]\} \quad (4)$$

where $0 \leq i \leq N_p - 1$, $\mathbf{A}^T$ denotes the transpose of matrix / vector $\mathbf{A}$.

Then a threshold $T_h\left(0 \leq T_h \leq N_p - 1\right)$ which is derived from the maximum channel delay spread is used to separate the time domain channel impulse response (4) into pure noise samples range / region and noise corrupted channel impulse response samples region. The samples in the pure noise region are used to estimate the noise variance via (5), i.e.,

$$\hat{\sigma}^2 = \frac{1}{N_p - T_h} \sum_{i=T_h}^{N_p-1} \left| \hat{h}_p[i] \right|^2. \quad (5)$$

Following that, the samples in the pure noise region are set zeros, i.e.,

$$\hat{h}_p[i] = 0, T_h \leq i \leq N_p - 1. \quad (6)$$

Then the estimated noise variance in (5) is used to remove the noise within the noise corrupted channel impulse response samples range, i.e.,

$$\hat{h}_p[i] = 0 \text{ if } \left|\hat{h}_p[i]\right|^2 < c\hat{\sigma}^2 \text{ for } 0 \leq i \leq T_h - 1, \quad (7)$$

where the constant $c$ is determined by certain optimization rule [7].

After that, the modified time domain channel impulse response in (6) and (7) is zero-padded, i.e.,

$$\hat{h}[l] = \begin{cases} \hat{h}_p[l], 0 \leq l \leq N_p - 1 \\ 0, N_p \leq l \leq N - 1 \end{cases}. \quad (8)$$

Finally, an $N$-point DFT is performed on the zero-padded channel impulse response in (8) to obtain the frequency domain channels corresponding to all the subcarriers of the $m$th OFDM symbol, i.e.,

$$\hat{H}[k] = DFT\left\{\hat{h}[l]\right\}, 0 \leq l, k \leq N - 1.$$

### III. THE IMPROVED DFT-BASED CHANNEL ESTIMATION

As we can see from last section, in conventional DFT-based channel estimation schemes, a *priori* channel information, i.e., the maximum channel delay spread, is needed, which is used to derive a threshold $T_h$ for separating the time domain channel impulse response samples. However, in wireless mobile communications, the channel varies from time to time, e.g., from one channel model to another, thus the maximum channel delay spread is changing. Besides, the maximum channel delay spread is not easy to be accurately estimated in practical scenarios. If the estimation of the maximum channel delay spread is inaccurate, the channel estimation performance might significantly decrease and become unacceptable. Moreover, the number of reference signals on an OFDM symbol might be limited. Therefore, the number of noise samples would be smaller if the number of reference signals on an OFDM symbol is small, which would lead to poor noise variance estimation causing poor channel estimation performance.

In the following, we will propose an improved DFT-based channel estimation scheme that utilizes the reference signals on multiple OFDM symbols to perform noise variance estimation which enlarges the number of noise samples. Besides, the proposed scheme does not need any *priori* channel information which makes it suitable for much wider applications.

The proposed algorithm is as follows. Firstly we transform the frequency domain channels corresponding to the reference signal subcarriers on all $M$ OFDM symbols, instead of one OFDM symbol like conventional schemes, into the time domain via IDFT, i.e.,

$$\hat{h}_p[i] = IDFT\left\{\left[\hat{\mathbf{H}}_p[0]; \cdots; \hat{\mathbf{H}}_p[m]; \cdots; \hat{\mathbf{H}}_p[M]\right]\right\}, \quad (9)$$

where $0 \leq i \leq N_p M - 1$.

Let's arrange $\hat{h}_p[i]$'s in (9) into a matrix as follows,

$$\hat{\mathbf{h}}_p = \begin{bmatrix} \hat{h}_p[0] & \cdots & \hat{h}_p[v] & \cdots & \hat{h}_p[M-1] \\ \vdots & \ddots & \vdots & \ddots & \vdots \\ \hat{h}_p[nM] & \cdots & \hat{h}_p[nM+v] & \cdots & \hat{h}_p[(n+1)M-1] \\ \vdots & \ddots & \vdots & \ddots & \vdots \\ \hat{h}_p[(N_p-1)M] & \cdots & \hat{h}_p[(N_p-1)M+v] & \cdots & \hat{h}_p[N_pM-1] \end{bmatrix}$$

$$= \begin{bmatrix} \hat{\mathbf{h}}_p[0] & \cdots & \hat{\mathbf{h}}_p[v] & \cdots & \hat{\mathbf{h}}_p[M-1] \end{bmatrix},$$

where $\hat{\mathbf{h}}_p[v]$ is the $v$th column of matrix $\hat{\mathbf{h}}_p$.

If the time duration of these $M$ OFDM symbols is much smaller than the channel coherence time, then the channels on different OFDM symbols of the same reference signal subcarrier index can be seen as the same. For example, on the $n$th reference signal subcarrier, we have $H_p[n, m] = C$ for $0 \leq m \leq M - 1$, where $C$ is a constant. Therefore, if there is no noise, all $\hat{\mathbf{h}}_p[v]$ for $1 \leq v \leq M - 1$ are zeros according to the property of DFT / IDFT. When there is noise, $\hat{\mathbf{h}}_p[v]$ for $1 \leq v \leq M - 1$ will be no longer zeros. As a matter of fact, $\hat{\mathbf{h}}_p[v]$ for $1 \leq v \leq M - 1$ are pure noise samples. Hence, we can use these pure noise samples to estimate noise variance as follows. Define

$$\hat{\mathbf{h}}_{p,\text{noise}} = \left[\hat{\mathbf{h}}_p^T[1] \cdots \hat{\mathbf{h}}_p^T[v] \cdots \hat{\mathbf{h}}_p^T[M-1]\right],$$

then we have

$$\hat{\sigma}^2 = \frac{1}{N_p(M-1)} \hat{\mathbf{h}}_{p,\text{noise}} \hat{\mathbf{h}}_{p,\text{noise}}^H, \quad (10)$$

where $\mathbf{A}^H$ denotes the Hermitian of matrix $\mathbf{A}$.

After obtaining the noise variance in (10), the noise in the time domain channel response $\hat{\mathbf{h}}_p[0]$ can be removed, i.e.,

If $\left|\hat{h}_p[nM]\right|^2 < \hat{\sigma}^2$, set $\hat{h}_p[nM] = 0$ for $n = 0:1: N_p - 1$. (11)

Then the cleaned time domain channel response in (11) is zero-padded, i.e.,

$$\hat{h}[l] = \begin{cases} \hat{h}_p[lM], 0 \leq l \leq N_p - 1 \\ 0, N_p \leq l \leq N - 1 \end{cases}$$

Finally, an $N$-point DFT is performed to obtain the frequency domain channels at all subcarriers,

$$\hat{H}[k] = DFT\left\{\hat{h}[l]\right\}, 0 \leq l, k \leq N - 1.$$

As we can see from the above procedure, the proposed scheme doesn't need any *priori* channel information, which is practical for channel estimation in wireless mobile communication systems. Besides, the noise variance estimation in the proposed scheme is more accurate than that in the conventional scheme since the number of noise samples is about several, e.g. $M - 1$, times larger, which takes advantage of the reference signals on multiple consecutive OFDM symbols.

### IV. SIMULATION RESULTS

In this section, we validate the proposed improved DFT-based channel estimation algorithm via simulations. The performance of the ideal channel estimation is shown as a benchmark. The simulations were performed on an OFDM-based simulator following the LTE framework. The number of subcarriers in an OFDM symbol was $N = 512$. The modulation type is QPSK. The reference signal subcarrier spacing is 8 subcarriers, therefore $N_p = 64$. The channel model is ETU5 defined in LTE [8]. The value of $M$ is 2. The signal-to-noise ratio (SNR) is defined as time-domain sample SNR.

In Fig. 2, the "Conv. CE, perfect" curve represents the performance of the conventional scheme when the estimation of the maximum channel delay spread is perfect. As we can see, it's acceptable and about 2.5 dB inferior to the performance of ideal channel estimation at 0.1% BER. We may wonder how the performance will be affected by the inaccurate estimation of the maximum channel delay spread in the conventional DFT-based channel estimation. Specifically, the estimated maximum channel delay spread is smaller than the real one, which leads to a smaller threshold $T_h$. Therefore, some real channel impulse response samples are separated as noise samples, which not only directly cause the reduction of the channel power but also overestimate the noise variance further leading to the removing of channel power in the process of denoising the samples within the maximum channel delay spread. In Fig. 2, the "Conv. CE inaccurate" curve represents the performance of the conventional scheme where the estimation of the maximum channel delay spread is inaccurate, e.g., the estimated channel delay spread is one tap smaller than the real one. As we can see from Fig. 2, the performance of the conventional scheme at high SNR is unacceptable when the estimation of the maximum channel delay spread is inaccurate. This is because the channel power is seriously reduced. In Fig. 2, the "Proposed CE" curve represents the performance of the proposed scheme. As we can see, the performance of the proposed scheme is closer to the performance of ideal channel estimation and about 1.5 dB better than the conventional scheme with ideal maximum channel delay spread estimation at 0.1% BER. The reason is the noise variance estimation in the proposed scheme is more accurate than that in the conventional scheme, which uses the reference signals on multiple consecutive OFDM symbols leading to more available noise samples.

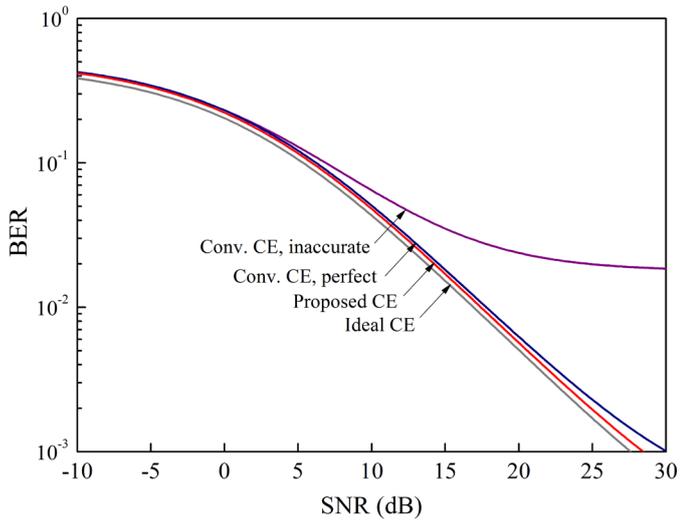

Fig. 2. Uncoded BER performance of different channel estimation algorithms.

## V. CONCLUSION

Channel estimation is an important part of mobile communication systems in realizing high data rate for emerging applications on smart devices. In this paper, we proposed an improved DFT-based channel estimation scheme which does not need any *priori* channel information (e.g., channel delay spread) compared to conventional schemes. Besides, the performance of the proposed scheme is better than that of the conventional scheme using perfect channel delay spread information. The reason is that the proposed scheme estimates the noise variance using the property of DFT via the reference signals on multiple consecutive OFDM symbols. By this way, the number of noise samples increases which naturally increases the accuracy of noise variance estimation. In a certain view, the scheme proposed in this paper is kind of soft without using any priori channel information while the conventional channel estimation schemes are kind of hard using *priori* channel information. The performance of the proposed scheme is further validated through simulations and compared with the conventional scheme.